\documentstyle[11pt,paspconf,epsf]{article}

\begin{document}

\title{High Energy Neutrinos from Blazars$^*$ 
\hspace{2.0cm} {\small ADP-AT-96-7}}
\author{R.J. Protheroe}
\affil{Department of Physics and Mathematical Physics,
The University of Adelaide, Adelaide, Australia 5005} 

\begin{abstract}
I describe a model of energetic gamma ray and neutrino emission 
in blazars which is consistent with most of the gamma-ray observations, and
use this model to estimate the diffuse intensity of high energy neutrinos
from blazars.\\ \vspace*{8mm}
{\small $^*$Accretion Phenomena and Related Outflows, IAU Colloq. 163, 1996} 
\end{abstract}


\vspace*{-15mm}

\section{Introduction}

The second  EGRET catalog of high-energy gamma-ray sources
(Thompson 1995) contains 40 high confidence identifications of
AGN, and all appear to be blazars.  
Clearly, the gamma ray emission is associated with the jet.  
Several of these AGN show gamma-ray variability with
time scales of $\sim 1$ day (Kniffen 1993), and this
places an important constraint on the models. 

Most previous work on gamma ray emission in AGN jets 
involved electron acceleration and inverse Compton scattering.
The proton blazar model was proposed originally by Mannheim and
Biermann (1992) and developed by Mannheim (1996).
In this model both protons and electrons are accelerated, and protons
interact via pion photoproduction and Bethe-Heitler pair production
with synchrotron radiation produced by the electrons.
Here I consider a model in which protons are accelerated
and interact directly with accretion disk radiation.

\section{The model}

If the protons interacted with matter to produce the gamma rays 
on a timescale consistent with the observed variability, 
then the proton number density in the blob must be 
$n_H > 10^9$ cm$^{-3}$, assuming a Doppler factor of $\sim 20$.
Such a high density would not normally be expected in AGN jets.
Interactions with the radiation field are therefore necessary if
protons are accelerated, but a sufficiently high
density of target photons for proton interactions would present problems 
for gamma ray escape if the radiation were isotropic.
In the present work, the gamma-ray emission region 
(referred to as ``the blob'')
is placed well above the accretion disk, and any scattering by 
clouds or coronae is assumed to be negligible, so that
the radiation impinging on the blob is highly anisotropic.
I use a spectrum which is a modification of the 
standard thin accretion disk
spectrum (Shakura and Sunyaev 1973)
in which the maximum temperature is lower than usual to account for the 
absence of the inner part of the disk due to the presence of the jet.
This is only an approximation, and does not include the soft X-ray 
emission.
Calculations are in progress using a more appropriate spectrum and
will be reported elsewhere.
However, the results for GeV gamma-rays and high energy neutrinos 
are not expected to differ by very much from those presented here.

From variability arguments, the size of the blob can not be
larger than $c \Delta t_{\rm var}'$ (primed quantities refer to the jet frame).
This means that the region in which the protons are accelerated
and trapped must be smaller than 0.02 pc, and proton gyroradii, $r'_g$,
at the highest energies must be less than 0.01 pc.
The maximum acceleration rate for any acceleration mechanism is
$(dE'/dt')_{\rm max} = ec^2B'$ where $B'$ is the magnetic field in the blob,
but for realistic acceleration mechanisms the acceleration rate will
be at least a factor of 10 lower.
During acceleration, protons will also suffer synchrotron losses 
and we require that these be less than the energy losses
for pion photoproduction at all energies 
up to the maximum energy (which is determined by pion photoproduction).
For a magnetic field of 10 gauss all these conditions are met,
giving a maximum jet-frame energy of 
$E_{\rm max}' \approx 1.5 \times 10^{10}$ GeV.

Protons may escape from the blob before interacting, and I assume this
is simply due to diffusion to the boundary of the blob where they
freely escape.
For a reasonable diffusion coefficient, most protons with energies below
the threshold for pion photoproduction will escape without interacting.

\section{Interaction of protons and the pair-synchrotron cascade}

We inject an $E'^{-2}$ spectrum of protons, continuously randomize
their directions in the jet frame, and allow them to interact with radiation.
The interactions are treated as described in Szabo and Protheroe (1994) and
all pions, and their decay products, are assumed to be produced travelling 
initially in the direction of the interacting proton.  
\begin{figure}[htb]
\plotone {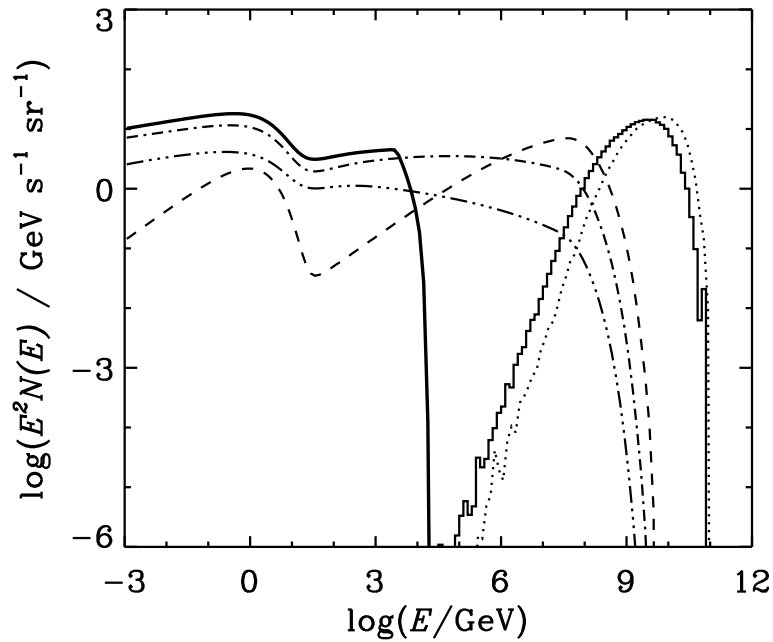}
\caption{Average gamma ray and neutrino spectra for viewing angles 
$2^\circ$ -- $4^\circ$ with respect to the jet axis resulting from 
$p\gamma$ interactions and subsequent cascading.  Full hisogram shows the total 
neutrino output ($\nu_\mu + \bar{\nu}_\mu + \nu_e + \bar{\nu}_e$), dotted 
hisogram shows the $\pi^0$ gamma spectrum on production, and dashed, 
dot-dashed, and dot-dot-dot-dashed curves show 1st, 2nd and 3rd generation 
synchrotron spectra on production.  Full curve shows the emerging gamma ray 
spectrum.}
\end{figure}

The gamma rays from $\pi^0$ decay may interact with photons from 
the accretion disk while inside the blob, or outside the blob, or
may escape from the AGN.
The probability of escaping from the blob 
and escaping from the AGN are calculated for various
angles to the jet axis.
The gamma rays that pair produce inside the blob will initiate a
pair-synchrotron cascade.
The electrons produced inside the blob by photon-photon pair production 
of gamma-rays from $\pi^0$ decay, and electrons from 
from $\pi^\pm$ decay, will synchrotron radiate in the ambient magnetic
field and produce the first generation of synchrotron 
photons in the cascade.
I then treat the photon-photon pair production interactions of this
first generation of synchrotron gamma-rays in the same way as 
gamma rays from $\pi^0$ decay, the resulting electron positron pairs
producing the second generation of synchrotron photons in the cascade.
This procedure is repeated for several generations.
The resulting gamma ray and neutrino fluxes emerging from the AGN
at angles between $2^\circ$ and $4^\circ$ with respect to the jet axis
are shown in Fig.~1.

\section{Neutrino background from unresolved blazars}

I use the luminosity function of gamma-ray loud AGN (Chiang 1995)
and integrate over redshift and luminosity
to estimate the diffuse background shown in Fig.~2.
Also shown is the result of Mannheim (1995) which, given the differences
in the models, is in surprisingly good agreement with the present work.
The reason for this is that despite the model differences, the most
important parameters have similar values (e.g. we inject the same proton 
spectrum, have similar maximum energies, similar magnetic fields,
and similar maximum soft photon energies).

The predicted neutrino intensities may be detectable with the
generation of neutrino telescopes currently being constructed
(see Gaisser et al. 1995 for a review).
For the present work,
the number of muon events per year in a detector with an effective 
collecting area of $2 \times 10^4$ m$^2$ 
would be $\sim 10$ above 1 TeV, and $\sim 6$ above 10 TeV, 
compared with the atmospheric neutrino background
of $\sim 225$ and 5 per year respectively (Hill 1996).

\begin{figure}[htb]
\plotone {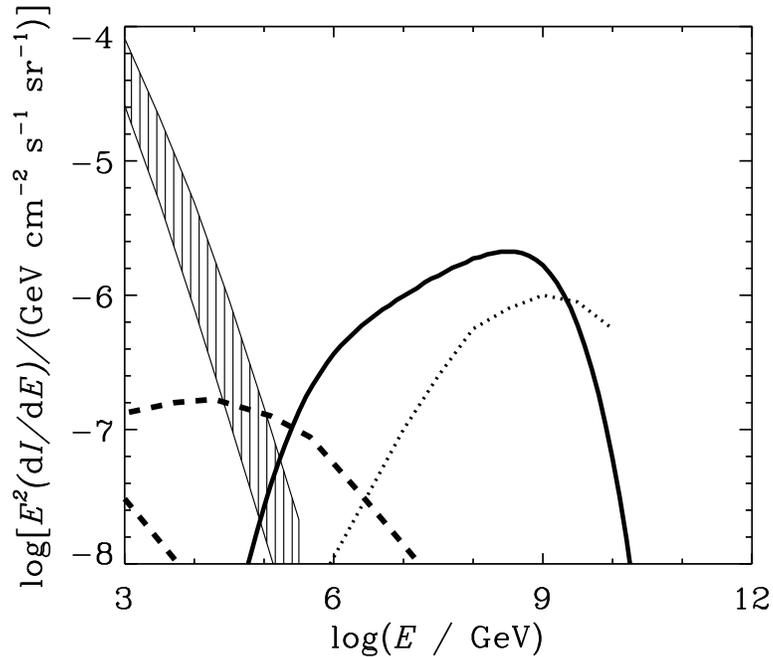}
\caption{The expected diffuse  
$\nu_{\mu} + \bar{\nu}_{\mu}$ intensity 
compared with the atmospheric neutrino intensity (Lipari 1993)
(vertical hatched band: the upper curve corresponds to zenith angle 
$\theta = 90^{\circ}$
and the lower curve corresponds to $\theta = 0^{\circ}$).
Also shown are possible minimum and maximum backgrounds due to charm
production (dashed lines).
Dotted curve shows the result of Mannheim (1995).}
\end{figure}

As a final note, I mention that one
of the EGRET blazars, Markarian 421, which was previously observed at
TeV energies (Punch et al., 1992), very recently had two violent 
outbursts at TeV energies
with a variability timescale of $\sim 1$ h (Gaidos et al. 1996).
Such rapid variability has yet to be adequately explained
by any of the models, including that described in the present work.

\end{document}